\documentclass[aps,pra,superscriptaddress,twocolumn,reprint]{revtex4-1}

\usepackage{verbatim} 
\usepackage{ulem} 
\usepackage{lipsum}
\usepackage{color}
\usepackage{amsmath}
\usepackage{graphicx}
\usepackage[caption=false]{subfig}
\usepackage{floatrow}

\renewcommand{\Re}{\operatorname{Re}}

\renewcommand{\Im}{\operatorname{Im}}

\begin{document}

\title{Linear and Non-linear Rabi Oscillations of a Two-Level System Resonantly Coupled to an Anderson-Localized Mode}

\author{Nicolas Bachelard}
\affiliation{ESPCI ParisTech, PSL Research University, CNRS, Institut Langevin, 1 rue Jussieu, F-75005, Paris, France}

\author{R\'{e}mi Carminati}
\affiliation{ESPCI ParisTech, PSL Research University, CNRS, Institut Langevin, 1 rue Jussieu, F-75005, Paris, France}

\author{Patrick Sebbah}
\affiliation{ESPCI ParisTech, PSL Research University, CNRS, Institut Langevin, 1 rue Jussieu, F-75005, Paris, France}
\affiliation{Department of Physics, The Jack and Pearl Resnick Institute for Advanced Technology, Bar-Ilan University, Ramat-Gan, 5290002 Israel}

\author{Christian Vanneste}
\affiliation{Laboratoire de Physique de la Mati\`ere Condens\'ee, CNRS UMR 3776,
Universit\'e de Nice-Sophia Antipolis, Parc Valrose 06108 Nice Cedex 02, France}
\email{christian.vanneste@unice.fr}

\date{\today}

\begin{abstract}
We use time-domain numerical simulations of a two-dimensional (2D) scattering system to study the interaction of a collection of emitters resonantly coupled to an Anderson-localized mode. 
For a small electric field intensity, we observe the strong coupling between the emitters and the mode, which is characterized by linear Rabi oscillations.
Remarkably, a larger intensity induces non-linear interaction between the emitters and the mode, referred to as the dynamical Stark effect, resulting in non-linear Rabi oscillations.
The transition between both regimes is observed and an analytical model is proposed which accurately describes our numerical observations.  
\end{abstract}


\maketitle

\section{Introduction}
The development of optical emitters at the micro and nanoscales, such as quantum dots, has allowed to investigate fundamental phenomena of light-matter interaction in practical experiments.
Among them, cavity quantum electrodynamics (cQED) is devoted to the problem of coupling atoms with the field of a resonant cavity. 
When the coupling is weak, the energy is irreversibly transferred from the emitters to the cavity and immediately radiates to the far field \cite{Purcell1946}.
In contrast, in the strong coupling regime, the energy is reversibly exchanged and quantum effects can be observed \cite{Jean-MichelGrard2003}.
For single emitters, such oscillating exchange of energy results in the so-called vacuum Rabi splitting (VRS) in the spectral domain of the joint frequency of the atomic transition and the cavity mode.
While the strong coupling regime was extensively investigated in the 1980's for a collection of two-level atoms in two-mirror cavities \cite{Haroche1992}, the experimental achievement of the VRS has remained a major challenge for many years.  
After its observation for a single atom in the 1990's, the genuine VRS was finally achieved with a single quantum dot in 2004 \cite{Yoshie2004}. 
To reach the strong coupling regime, the exchange rate, \textit{i.e.} the Rabi frequency, must be larger than the mean of the atomic and the cavity decay rates. 
This condition is usually difficult to fulfil experimentally with a single quantum emitter because of the weakness of the single-photon field. 
This is why the first observations of Rabi splitting were realized with a collection of emitters rather than a single one \cite{Kaluzny1983,Weisbuch1992a}. 
Unlike the single-emitter strong coupling which is fundamentally quantum, the Rabi splitting obtained for many emitters is semiclassical \cite{Agarwal1984}.
In the literature, this semiclassical regime is referred to as non-perturbative normal-mode coupling (NPNMC) to make the distinction with true mode coupling \cite{Khitrova2006}. 
NPNMC has been achieved with different types of resonators (e.g. photonic crystal defect, micropillar, microdisk cavities \cite{Weisbuch1992a}).

A collection of two-level atoms resonantly excited by an intense electric field is another interesting example of light-matter interaction.  
In the early 1970's, Mollow \cite{Mollow1972} derived the rate absorption of two-level atoms driven by a strong excitation field.
He predicted non-linear atomic response resulting in stimulated emission in the absence of population inversion.
His prediction was confirmed experimentally several years later \cite{Ezekiel1977}.
From a physical point of view, the intense electric field is responsible for oscillations of the atomic population.
Such oscillations induce Stark shifts in the energy-level structure of the atoms leading to new resonances and therefore to a non-linear susceptibility.
This effect is also known as the Dynamic Stark Effect (DSE) \cite{Meystre2007,Boyd2008}. 
Emission resulting from DSE should not be confused with the resonance fluorescence of a two-level atom also introduced by Mollow \cite{Mollow1969}.
Even if both effects are closely related \cite{Meystre2007}, DSE can be describe semiclassically and is based on stimulated emission while resonance fluorescence relies on spontaneous emission.  

In disordered media the modal confinement may be solely driven by the scattering strength: When the scattering becomes sufficiently strong, the spatial extension of the modes is reduced and eventually results in modal confinement by disorder \cite{Anderson1958}.
Similarly to conventional cavities, Anderson-localized modes have demonstrated remarkable abilities to support different physical effects (e.g. mode coupling \cite{Labonte2012,Bachelard2014a} or lasing amplification \cite{Milner2005,Yang2011,Liu2014}).
Among them, NPNMC was recently achieved between localized modes and nano-emitters \cite{Sapienza2010,Thyrrestrup2012,Gao2013,Caze2013}, opening a new paradigm in the field of cQED.
In contrast, DSE was proposed as a coherent amplification mechanism in cold-atom clouds \cite{Guerin2008}.
Nevertheless, in such disordered systems, since atoms provide for the amplifcation their scattering strengh is very weak.  
Therefore, this raises two main questions.
First, can DSE be supported by the modes of a strongly disordered system, similarly to NPNMC?
Second, can DSE and NPNMC coexist in a single Anderson-localized mode? 

In this article, we investigate the dynamic competition between NPNMC and DSE resulting from the coupling between an Anderson-localized mode and a collection of two-level atoms.
We demonstrate that both effects lead to Rabi oscillations but of different kinds, namely the linear and non-linear Rabi oscillations. 
Time-domain numerical simulations are run to observe both regimes.
An analytical model is proposed to describe each regime and the transition from non-linear to linear oscillations.
This work confirms that the strong coupling between a single emitter and a 2D Anderson-localized mode investigated in \cite{Caze2013} can be observed for a collection of atoms.  
It also support the idea that Anderson-localized modes, similarly to conventional cavities, may serve as an interesting platform to study non-linear effects resulting from strong energetic confinement.
This paper is organized as follows.
In Sec.~\ref{SecII}, we use exact numerical simulations in the time domain to study the interaction between a spatially confined population of two-level atoms and a single Anderson-localized mode in a 2D disordered medium.
All the emitters are initially at ground state and the mode is excited by a monochromatic light pulse which induces oscillations of the atomic population inversion.
For large intensity, the population inversion saturates and gives rise to DSE splitting of the electric field in the frequency domain.
This regime is referred to as the non-linear Rabi regime.
For smaller intensity, the population difference no-longer saturates.
The oscillations of the population inversion decrease and slow down: The NPNMC regime between the collection of atoms and the mode is reached.
This regime is referred to as the linear Rabi regime.  
In Sec.~\ref{SecIII} and \ref{SecIV}, we investigate theoretically the non-linear and linear regimes of Rabi oscillations, respectively.
We derive analytically the splitting amplitude and the condition to reach both regimes.
In Sec.~\ref{SecV}, we analytically explain non-linear and linear oscillations observed in Sec.~\ref{SecII} and describe the transition observed between these two regimes. 
In Sec.~\ref{SecVI}, we summarize and emphasize potential applications of this work.

\section{Numerical investigation\label{SecII}}
In this section, we use exact numerical simulations in the time domain to highlight the existence of the different regimes of oscillations. 
A collection of two-level atoms is introduced in a 2D disordered system, where an Anderson-localized mode exist. 
An external source excites the localized mode. 
The resulting large intensity of the mode induces non-linear Rabi oscillations which are characterized by simultaneous oscillations of the polarization and of the population inversion.
When the excitation is turned off, the amplitude of the mode decays leading to a decline of the field at the position of the atoms. 
The oscillations of the population inversion decays and eventually vanish to give way to the strong coupling regime \textit{i.e} to linear Rabi oscillations of the polarization and of the field in quadrature.

\subsection{Description of the Model\label{sec:model}}
\begin{figure}[h!]
   \sidesubfloat[]{\includegraphics[scale=.585]{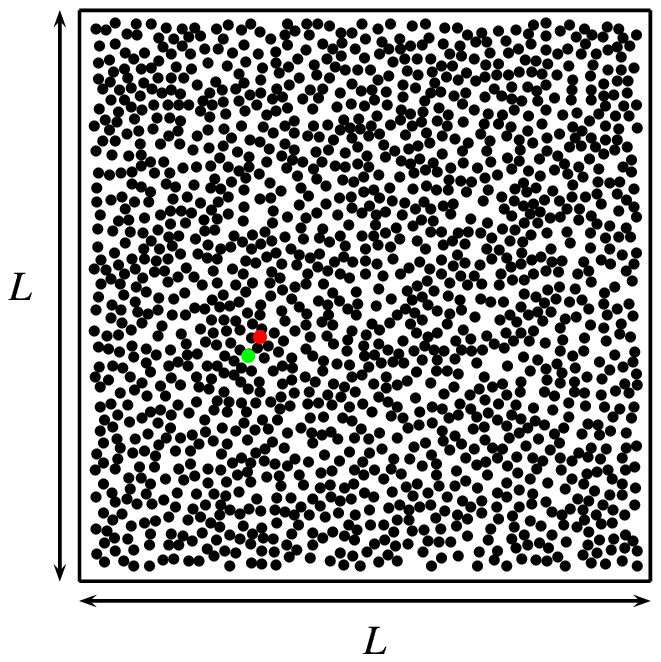}} \hfill
   \sidesubfloat[]{\includegraphics[scale=.275]{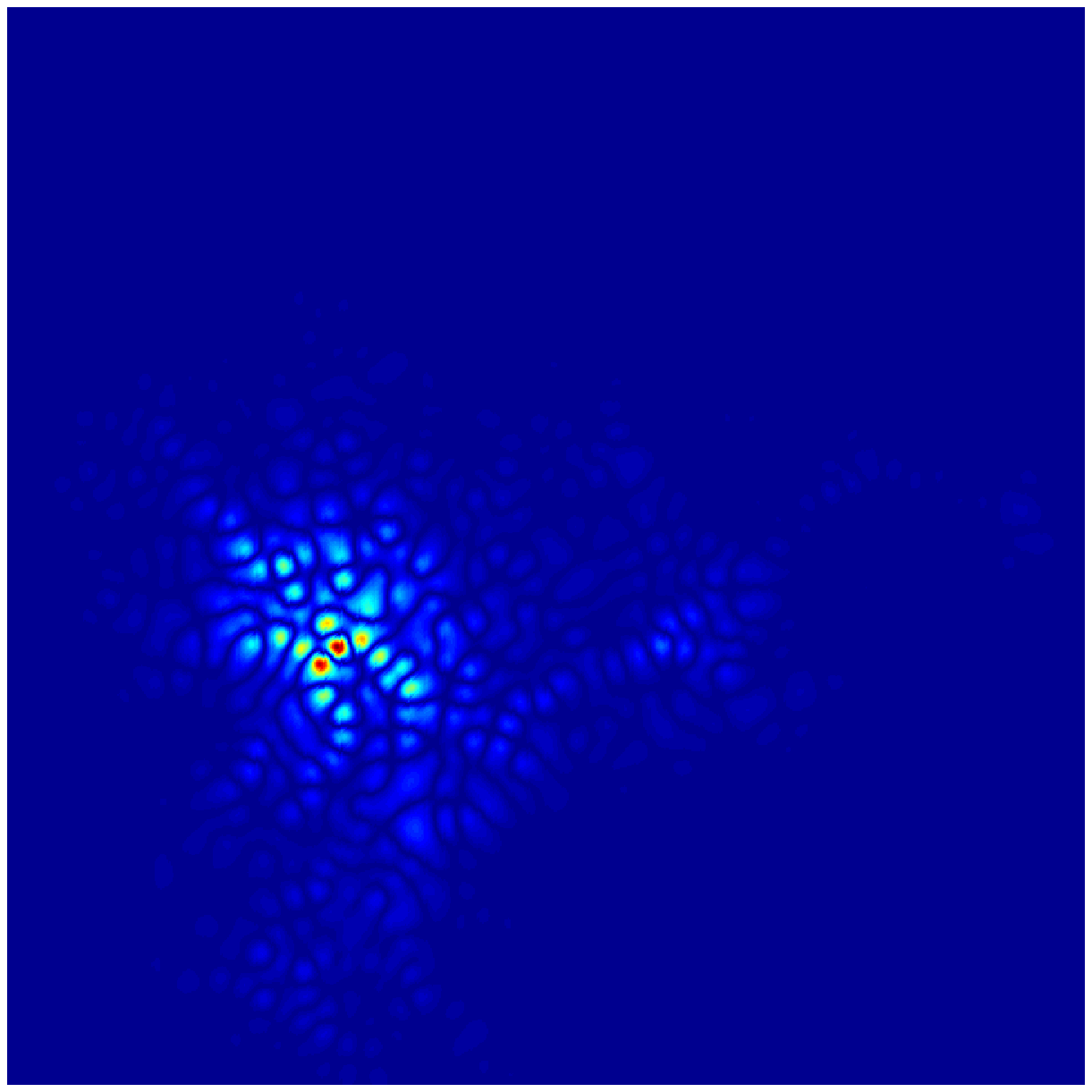}} \hfill
   \sidesubfloat[]{\includegraphics[scale=1]{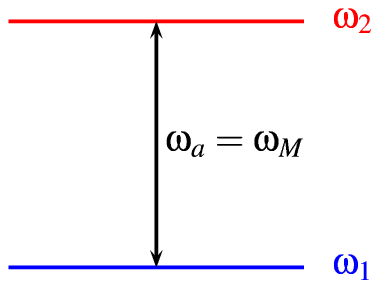}} 
\caption{(a) 2D disordered system composed of a randomly distributed ensemble of circular scatterers (radius $r = 60$ $nm$ and index $n = 2$) embedded in vacuum. System area size is $L^2 = 6.6$ $\mu m^2$. An ensemble of two-level-atom dipoles is inserted at position $\textbf{r}_{a}$ (\textcolor{red}{red dot}) and a point-like source is located at $\textbf{r}_{ext}$ (\textcolor{green}{green dot}). 
(b) Field spatial distribution of an Anderson-localized mode $M$ at $\lambda_M = 453.68$ $nm$, namely $\Psi_M(\textbf{r})$. 
(c) Energy-level structure of a two-level atom. Each atom is composed of a ground state of energy $E_1 = \hbar \omega_1$ and an excited state of energy $E_2 = \hbar \omega_2$. 
The transition between the two states, namely $\omega_a$, is resonant with mode $M$ frequency, $\omega_M = \omega_a$. 
} 
\label{Fig1}
\end{figure}

We consider a 2D random collection of circular dielectric particles 
with radius $r=60\, nm$ and optical index $n = 2$ embedded in a background medium of index $n=1$ (see Fig.~\ref{Fig1}(a)). 
The volume fraction is $40\,\%$ and the system size is $ L^{2} = 6.6 \times 6.6 \, \mu m^{2}$.
With the above choice of parameters, the system is known to be in the localized regime \cite{Vanneste2001,Sebbah2002}.
The dipole which will interact with one of the localized modes of the system corresponds to a collection of two-level atoms located at one node of the numerical system (see Fig.~\ref{Fig1}(a) and \ref{Fig1}(c)). 
Its density is given by $N/dx^{2}$ where $N$ is the total number of atoms and $dx$ is the space increment of the numerical grid. 
Using the notations of \cite{Vanneste2001,Sebbah2002}, the corresponding population equations read
\begin{align}
  dN_1/dt=& N_2/\tau_{21}-(\textbf{E}/\hbar\omega_a)\cdot d\textbf{P}/dt\nonumber\\
  dN_2/dt=& -N_2/\tau_{21}+(\textbf{E}/\hbar\omega_a)\cdot d\textbf{P}/dt, \label{Eq1}
\end{align}
where $N_i,   i=1,2$ are the populations of the two levels. 

\floatsetup[figure]{style=plain,subcapbesideposition=top}
The angular frequency of the atomic transition between levels 1 and 2 is $\omega_{a}=2\pi\nu_{a}$ (see Fig.~\ref{Fig1}(c)). 
$\textbf{E}$ and $\textbf{P}$ are the electric field and the polarization density, respectively.
Electrons in level 2 can decay to level 1 either spontaneously with time constant $\tau_{21} = $ 1 $ns$ or through 
stimulated emission with the rate $(\textbf{E}/\hbar\omega_a)\cdot d\textbf{P}/dt$. 
Conversely, electrons in level 1 can jump to level 2 with the absorption rate $(\textbf{E}/\hbar\omega_a)\cdot d\textbf{P}/dt$. 
The polarization obeys the equation
\begin{equation}
  d^2 \textbf{P}/dt^2 + \Delta \omega _a d\textbf{P}/dt +
  \omega _a^2 \textbf{P}=\kappa.\Delta N.\textbf{E}\label{Eq2}
\end{equation}
where $\Delta N=N_1-N_2$ is the population inversion. 
The linewidth of the atomic transition is given by the relaxation rate of the polarization $\Delta\omega_a = 1/\tau_{21} + 2/T_{\Phi}$, where $T_{\Phi}$ = 1 $ns$ is the decoherence time of the collection of atoms.
The constant $\kappa$ is given by $\kappa=6\pi\varepsilon_0c^3/(\omega^2_a\tau_{21})$ \cite{siegman1986lasers}. 
Finally, the polarization is a source term in the Maxwell equations
 \begin{align}
  \mu_0\partial\textbf{H}/\partial t=& -\nabla\times\textbf{E}\nonumber\\
  \varepsilon_0 \varepsilon(\textbf{r})\partial\textbf{E}/\partial t=&
  \nabla\times\textbf{H}-\partial\textbf{P}/\partial t\label{Eq3}
\end{align}
where $H$ stands for the magnetic field, $\varepsilon(\textbf{r}) = n^2(\textbf{r})$ the dielectric function of the scatterers assumed non-dispersive, $\mu_0$ and $\varepsilon_0$ the vacuum permeability and permittivity, respectively. 

\begin{figure}
\begin{center}
\includegraphics[clip,width=1.0\linewidth]{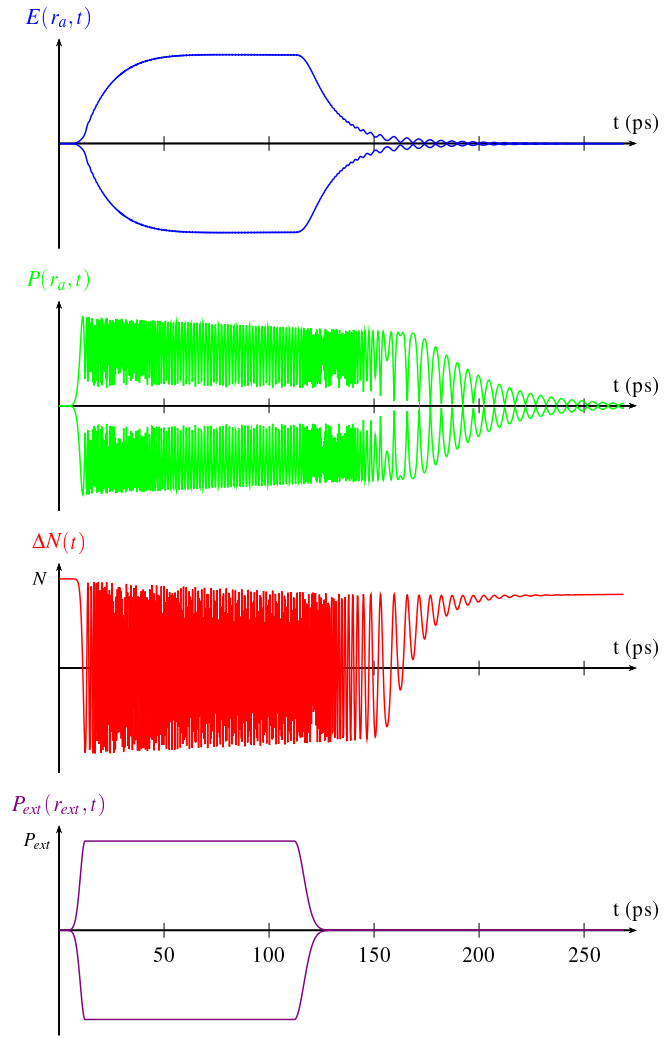}
\caption{Time evolution of the field recorded at the dipole position (blue curve), of the polarization (green curve) and of the population inversion (red curve). 
For a better clarity, oscillations at the optical frequency of the field and of the polarization are not displayed. 
Actually, the blue and green curves are envelopes of the time evolution of the field and the polarization. 
The purple line is the envelope of the source. 
The vertical units are arbitrary since the different curves have been rescaled in order to share the same vertical scale. 
Similar vertical rescaling have been applied to all figures which display the time evolution of the field, the polarization and the population inversion.} \label{Fig2}
\end{center}
\end{figure}
 
This set of equations is equivalent to the Maxwell-Bloch equations that are commonly used in standard semiclassical laser theory \cite{Milonni2010}. 
We have considered a 2D transverse magnetic field so that the electric field has only one component orthogonal to the plane of the 2D system.
Maxwell's equations are solved using the finite-difference time-domain (FDTD) method and C-PML absorbing boundary conditions are used in order to model an open system \cite{Taflove2005}.

\subsection{Numerical results\label{sec:numres}}
We carried out the numerical study in the following way. 
First, we study the spectrum and the modes of a random collection of scatterers according to the procedure which is described in detail in \cite{Sebbah2002}.
Among the modes, we deliberately choose a localized mode with a good quality factor and well isolated spectrally from the others in order to facilitate the strong coupling. 
An example of such a mode (referred to in the text as mode $M$) is displayed in Fig.~\ref{Fig1}(b). 
 
The frequency $\nu_{M}$ of this mode (corresponding to the wavelength $\lambda_{M} = 453.68 \, nm$) is known from the spectrum of the impulse response of the random system (not shown). 
Though this mode is well located inside the random system, there exists some leakage through the system boundaries. 
Hence if this mode oscillates freely without external excitation, its amplitude decays as a function of time. 
For this mode, the decay time is about $15 \, ps$, which corresponds to a quality factor $Q_M = 2.9\times 10^{4}$.

In order to improve the coupling between the mode and the dipoles, the next step is to introduce the collection of atoms at one node of the numerical system where the amplitude of the localized mode $M$ is large as shown in Fig.~\ref{Fig1}(a) and (b) (referred to in the text as position $\textbf{r}_a$). 
Moreover, the atomic frequency $\nu_{a}$ is deliberately set at the value $\nu_{M}$ of the localized mode. 
In other words, we have chosen $\omega _a =2\pi\nu_{M}$ in eq.~(\ref{Eq2}).

Finally, a monochromatic point source excites the system at position $\textbf{r}_{ext}$ in the vicinity of the dipoles (see Fig.~\ref{Fig1}(a)). 
Its frequency $\omega_{ext}$ matches the joint frequency of the atomic transition, $\nu_{a}$, and of the localized mode, $\nu_{M}$ . 
The duration $T_{ext} = 100\, ps$ of the source and of its gaussian transients is sufficient to avoid the excitation of the other modes of the system which are spectrally close (the envelope of the excitation pulse $P_{ext}(\textbf{r}_{ext},t)$ is displayed in Fig.~\ref{Fig2}). 
This results in the simultaneous excitation of the localized mode $M$ and of the atomic dipoles as shown by the time evolution of the field, the polarization and the populations inversion recorded at the position of the dipoles (Fig.~\ref{Fig2}).

\begin{figure}[ht!]
\begin{center}
   \sidesubfloat[]{\includegraphics[scale=.6]{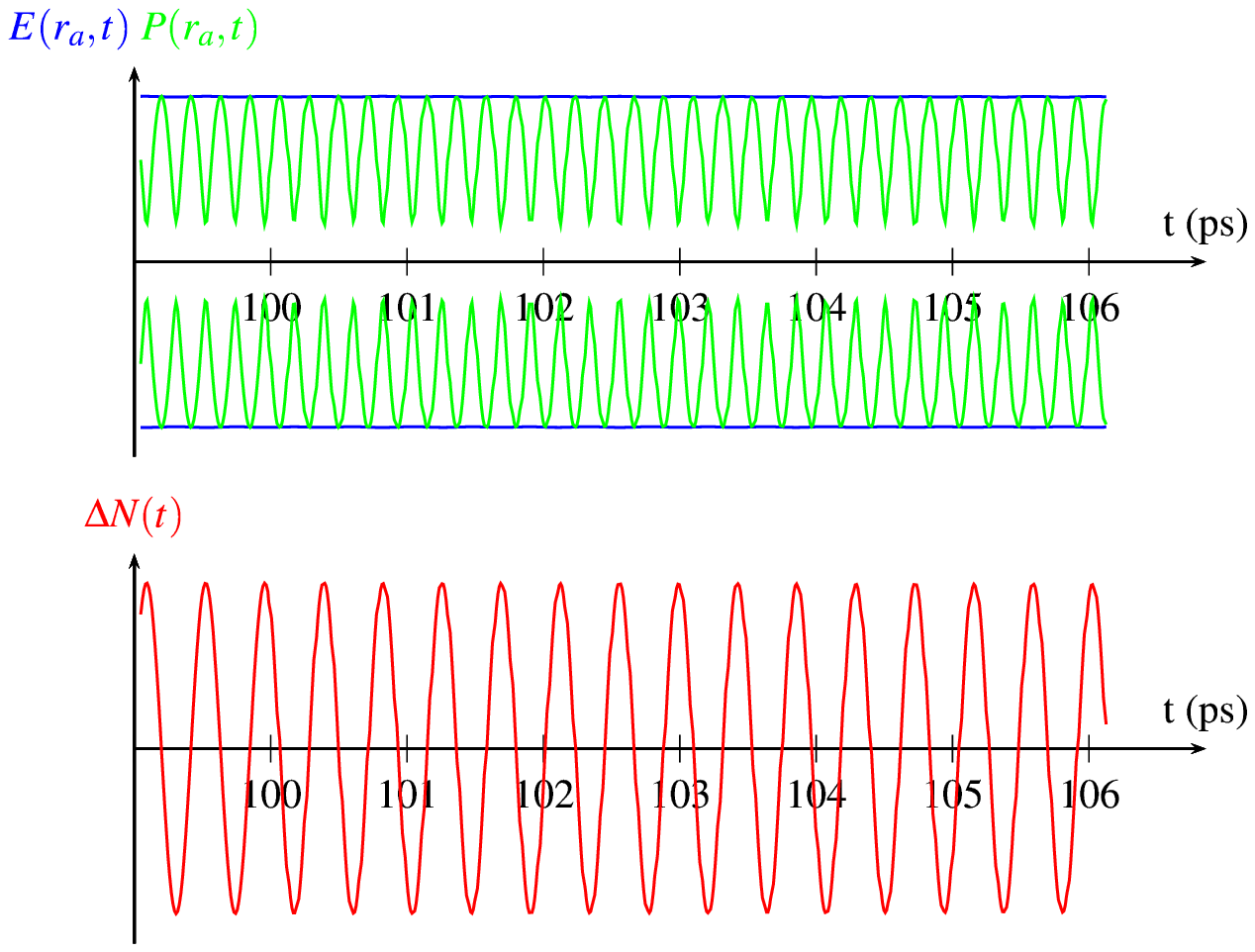}} \hfill
   \sidesubfloat[]{\includegraphics[scale=.6]{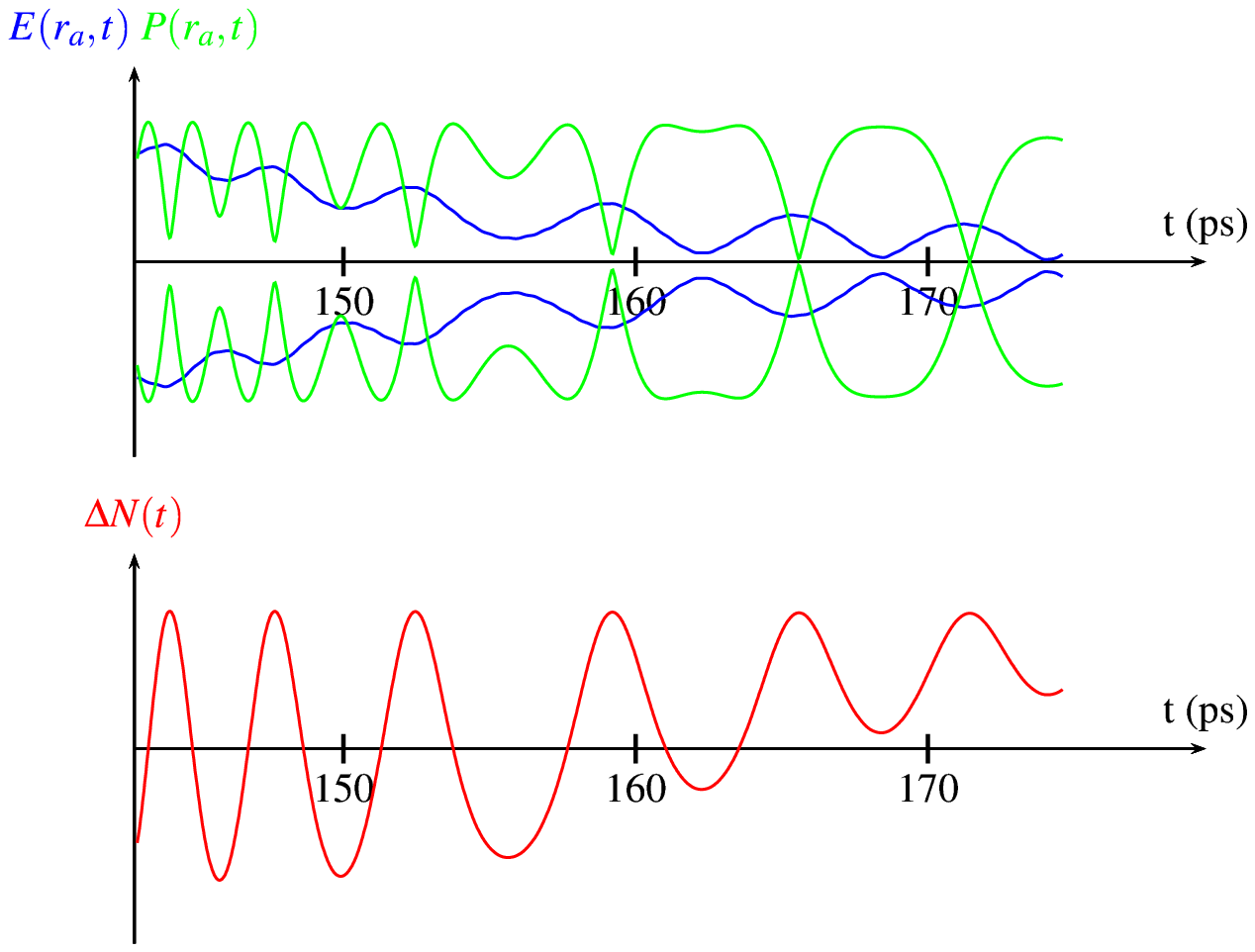}} \hfill
   \sidesubfloat[]{\includegraphics[scale=.6]{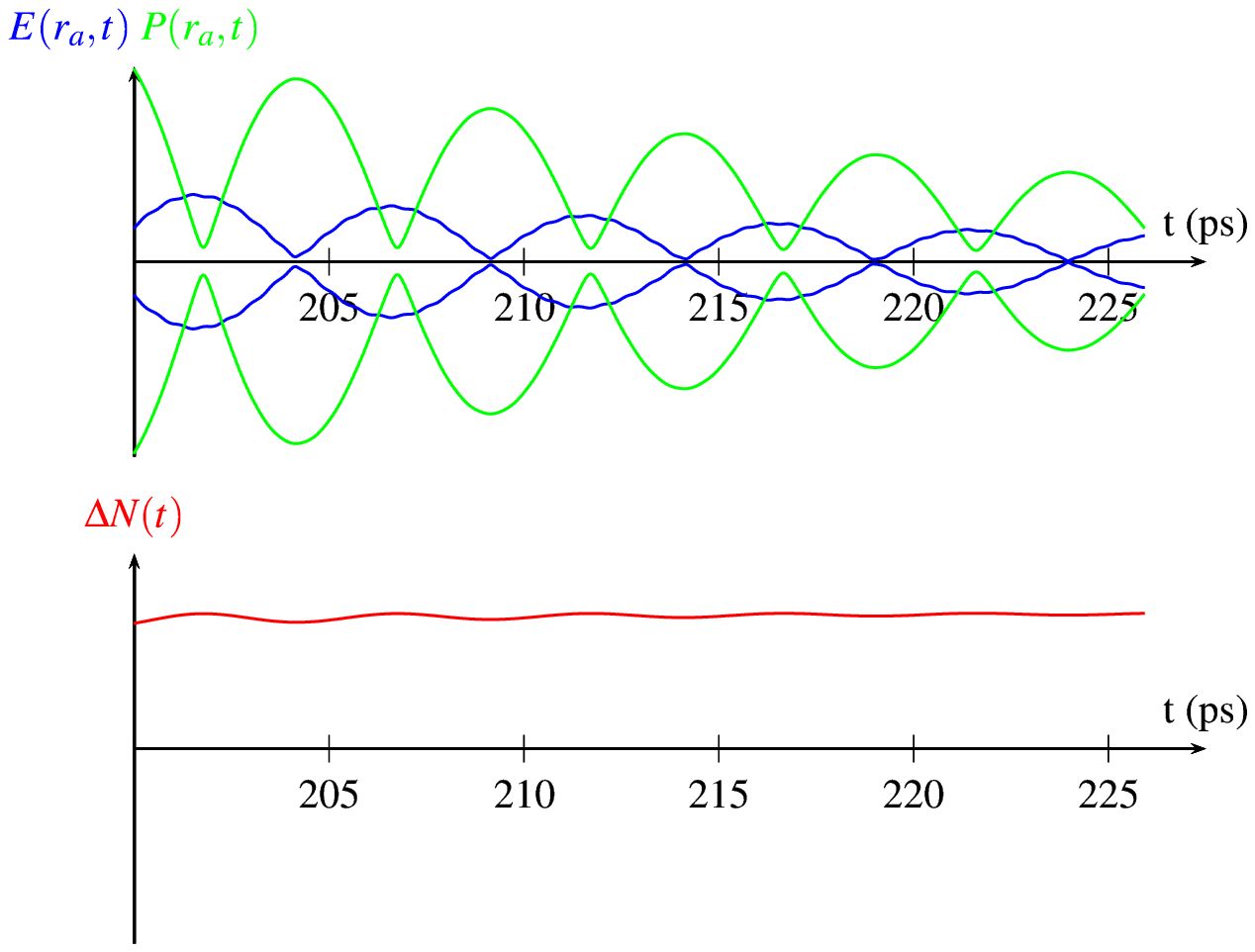}} 
\caption{Time evolution of the field (blue curve), the polarization (green curve) and population inversion (red curve) recorded at the dipole position, $\textbf{r}_a$, for three different intervals, $[99-106]$ $ps$ (a), $[143-174]$ $ps$ (b) and $[200-226]$ $ps$ (c). 
As in Fig.~\ref{Fig2}, the blue and green curves are envelopes of the time evolution of the field and of the polarization.}
\label{Fig3}
\end{center}
\end{figure}

At times earlier than $t\simeq 110 \, ps$, when the source is still on, the electric field $E(\textbf{r}_a,t)$ is almost monochromatic.
Such monochromatic excitation of the atoms results in oscillations of the population inversion, $\Delta N(t)$, and of the envelope of the polarization, $P(\textbf{r}_a,t)$.
The modulation of the envelope of the polarization $P(\textbf{r}_a,t)$ for a constant amplitude of the electric field indicates that the response of the atoms is non-linear. 
Oscillations are referred to as non-linear Rabi oscillations. 
After the external source is progressively turned off, starting from time $t\simeq 110\, ps$, a transient regime takes place as seen in Fig.~\ref{Fig2}, where the amplitudes of the field and of the polarization slowly decay due to the field leakage through the system boundaries. 
Correspondingly, oscillation frequency of the envelope of the polarization and the population inversion decrease until these oscillations vanish. 
At the end of the transient regime after $t \simeq 200 \, ps$, the population inversion is almost completely stabilized. 
Then, one observes Rabi oscillations of the field amplitude and of the polarization in quadrature (see Fig.~\ref{Fig3}(c)).
This regime is referred to as linear since the oscillations of both envelopes are similar to the frequency beating originating from two coupled oscillators. 

For a better illustration of the three regimes mentioned above, enlargements of the time evolution of the field, the polarization and the population inversion are shown in Fig.~\ref{Fig3} at three different time intervals. 

\begin{figure}
\begin{center}
\sidesubfloat[]{\includegraphics[clip,width=1.0\linewidth]{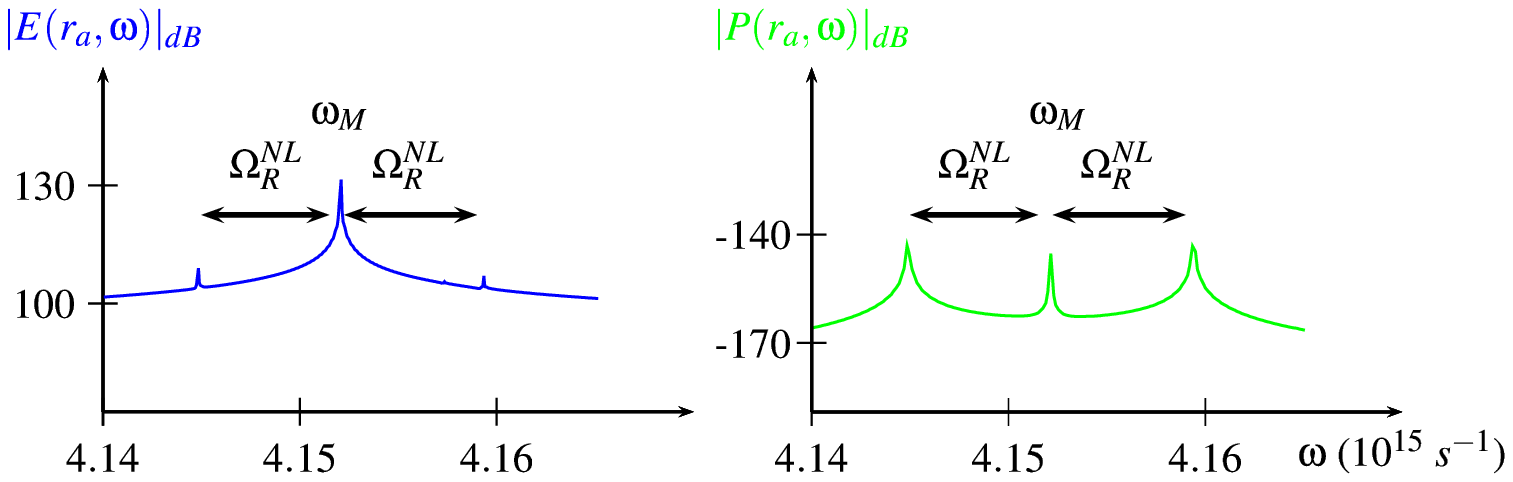}} \hfill
 \sidesubfloat[]{\includegraphics[clip,width=1.0\linewidth]{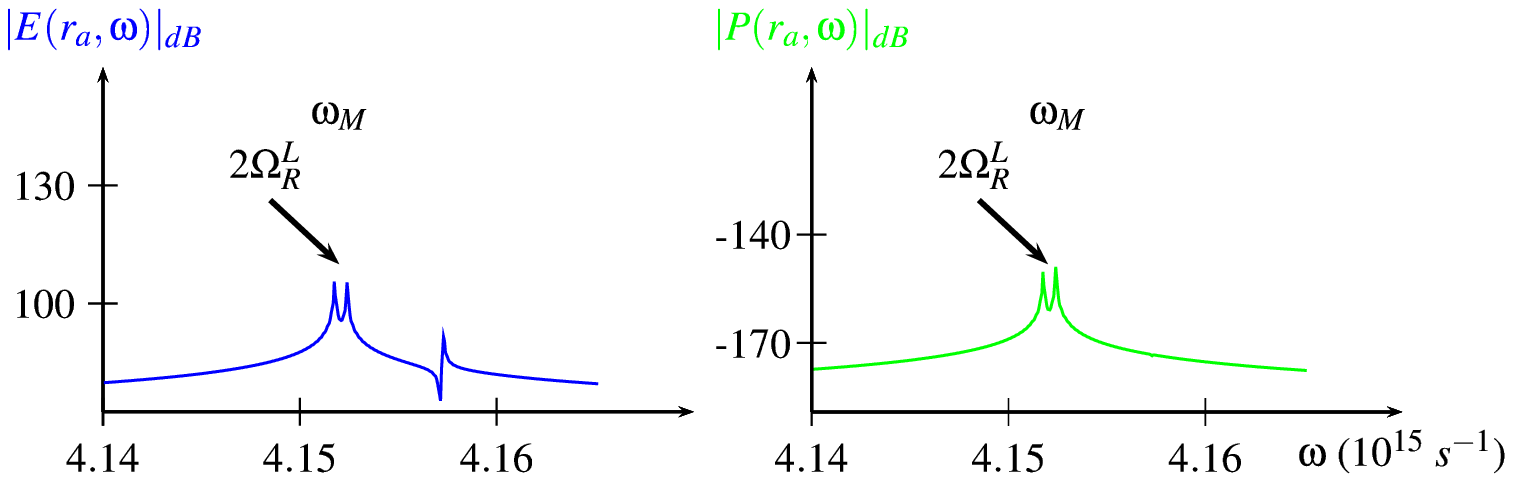}}
\caption{
(a) Spectra of the electric field (blue curve) and the polarization (green curve) in the time interval $[99-106] \, ps$. 
Both spectra exhibit three components but the contribution of sidebands on the electric field is weak. 
The electric field is almost monochromatic at frequency $\omega_M$.   
(b) Spectra of the electric field (blue curve) and the polarization density (green curve) in the time interval $[200-226] \, ps$. 
Both spectra exhibit similar splitting around the resonant frequency $\omega_a = \omega_M$. 
The presence of a third peak in electric field at $\approx $ 4.16 10$^{15} \, s^{-1}$ corresponds to another Anderson-localized mode initially excited by the monochromatic source. 
}
\label{Fig4}
\end{center}
\end{figure}

In the first time interval $[99-106] \, ps$, the source is active (purple curve in Fig.~\ref{Fig2}). 
While the field intensity displays a fixed dc level, one observes non-sinusoidal oscillations of the polarization which are associated with strong oscillations of the population inversion. 
The strong oscillations of $\Delta N$ are Rabi oscillations that are induced by the large amplitude of the field. 
The corresponding spectra of the electric field and polarization are displayed in Fig.~\ref{Fig4}(a) and exhibit three peaks instead of a single peak at $\omega_{M}= 2\pi c/\lambda_M \approx$ 4.15  $10^{15} \, s^{-1}$.

In the second time interval $[143-174] \, ps$, after the excitation pulse has been turned off, the amplitude of the field has significantly decreased due to leakage through the open boundaries. 
The frequency of the Rabi oscillations has also significantly decreased. 
Moreover starting at time $t\simeq140 \, ps$, their amplitudes decrease, especially the oscillations of the population inversion.

The third time interval $[200-226] \, ps$ has been chosen at a later time when the field amplitude is much smaller and the population inversion has almost stopped oscillating.
One observes beating oscillations of the field amplitude and of the polarization. 
The corresponding spectra are displayed in Fig.~\ref{Fig4}(b). 
The peak at $\omega_M \approx 4.15$ $10^{15} \, s^{-1}$ has been split in two separate peaks and characterize the linear Rabi regime in the frequency domain.
The oscillations of the field amplitude and of the polarization and the splitting of the initial peak constitute the hallmark of the strong coupling of the atoms and the localized mode. 
Note the almost symmetrical spectrum on either part of the initial peak. 
Also the field and the polarization oscillate in quadrature as expected for the regime of strong coupling \cite{Haroche1992}. 
These results are in agreement with recent studies of strong coupling with Anderson localize modes in the frequency domain \cite{Thyrrestrup2012,Caze2013}. 
The simulations presented here provide a complementary time domain description.

\section{Theoretical investigation of the Non-linear Rabi regime}\label{SecIII}
In this section, we describe analytically the regime of oscillation observed in time interval $[99-106] \, ps$ and referred to as the non-linear Rabi regime.
We derive the pulsation of non-linear Rabi oscillations and the observation conditions of this regime.

\subsection{Non-linear Rabi pulsation} 
\begin{figure}[ht!]
\begin{center}
\includegraphics[clip,width=1.0\linewidth]{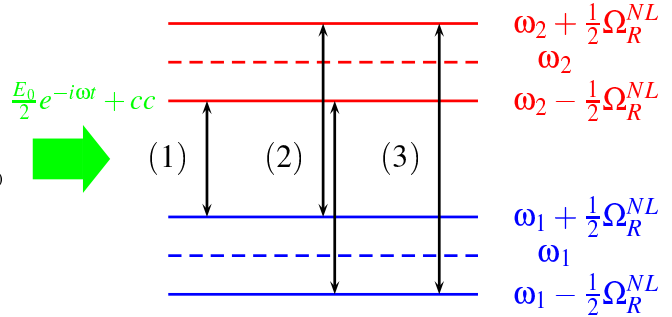}
\caption{
Schematic representation of the Dynamical Stark Effect: An atom in vacuum is described by two levels of energy, $E_{i\in[1,2]} = \hbar \omega_{i\in[1,2]}$, represented by the dashed lines and corresponding to a single transition. 
When an electric field, $\frac{E_0}{2}e^{-i\omega t}+cc$, is applied the population of each energy level oscillates. 
This results in a splitting of the energy levels $E_{i\in[1,2]} \pm \frac{1}{2} \hbar \Omega_R^{NL}$ and three different transitions. 
} 
\label{Fig6}
\end{center}
\end{figure}

When its intensity is large, the electric field saturates the collection of two-level atoms and substantially changes their energy-level structure. 
This saturation induces Stark shifts on energy levels corresponding to oscillations of the population inversion of pulsation $\Omega_R^{NL}$, referred to as non-linear Rabi pulsation (see Fig.~\ref{Fig3}(a)).
As sketched in Fig.~\ref{Fig6}, energy levels $E_{i\in[1,2]} = \hbar \omega_i$ are split into $E_{i\in[1,2]} \pm \frac{1}{2} \hbar \Omega_R^{NL}$.
The occurrence of Stark shifts is responsible for new resonances and atoms are described by a non-linear optical susceptibility.
The non-linear susceptibility induces sideband frequencies in the polarization and the electric field at $\omega \pm  \Omega_R^{NL}$ (see Fig.~\ref{Fig4}).

The non-linear Rabi pulsation $\Omega_R^{NL}$ can be derived from different ways (e.g. dressed states \cite{Autler1955,Boyd1988} or bare states \cite{Lamb1964,Sargent1978,Boyd1988}).
Here, we consider an approach based on rate equation \cite{siegman1986lasers}.

To saturate the two-level atoms, the intensity of the field must be very large at the position of the atoms, namely $\textbf{r}_a$.
As illustrated in Fig.~\ref{Fig3}(a) and Fig.~\ref{Fig4}(a), the sideband contribution is therefore weak and the electric field is almost monochromatic
\begin{equation}
   E(\textbf{r}_a,t) \approx \frac{E_0}{2} e^{-i\omega t} + cc
   \label{Eq4}
\end{equation} 
In contrast, non-linear susceptibility of the atoms deeply affects the polarization, which reads
\begin{equation}
   P(\textbf{r}_a,t) =  p(t) e^{-i\omega t} 
   \label{Eq5}
\end{equation} 
where $p(t)$ stands for the envelope amplitude.
The envelope of the polarization is assumed to vary slowly with respect to the optical frequency: $\omega^2|p(t)| \gg \omega |dp(t)/dt|  \gg | d^2p(t)/dt^2|$.
Under this assumption, the second derivative of the polarization can be derived from eq.~\eqref{Eq5}
\begin{equation}
   \frac{d^2P}{dt^2}(\textbf{r}_a,t) \approx -2 i \omega \frac{dP}{dt}(\textbf{r}_a,t) + \omega^2 P(\textbf{r}_a,t) 
   \label{Eq5bis}
\end{equation} 
The evolution of the polarization, driven by eq.~\eqref{Eq2}, can be simplified using eq.~\eqref{Eq5bis} and the rotation wave approximation ($(\omega_a - \omega)^2 \approx 2 \omega_a (\omega- \omega_a)$)
\begin{equation}
   \frac{dP(\textbf{r}_a,t)}{dt} \approx  -\left(i\omega_a + \frac{\Delta \omega_a}{2}\right) P(\textbf{r}_a,t) + i \frac{\kappa}{2 \omega_a} \Delta N(t) E(\textbf{r}_a,t) 
   \label{Eq6}
\end{equation} 
Injecting the expression of the electric field given by eq.~\eqref{Eq4} and neglecting the non-resonant terms, eq.~\eqref{Eq6} reads
\begin{equation}
   \frac{dp(t)}{dt} =  \left(i(\omega - \omega_a) - \frac{\Delta \omega_a}{2}\right) p(t) + i \frac{\kappa}{4 \omega_a} \Delta N(t) E_0 
   \label{Eq7}
\end{equation} 
From eq.~\eqref{Eq1} we derive the population inversion evolution
\begin{equation}
   \frac{d\Delta N(t)}{dt} =  2 \frac{N_2}{\tau_{21}} - \frac{1}{2 \hbar \omega_a} \left(E_0 e^{-i\omega t}+cc \right) \left(\frac{dP(\textbf{r}_a,t)}{dt}+cc\right)
   \label{Eq8bis}
\end{equation} 
For a slow envelope, eq.~\eqref{Eq8bis} can be simplified  
\begin{equation}
   \frac{d\Delta N(t)}{dt} =  \frac{N}{\tau_{21}} - \frac{\Delta N}{\tau_{21}} - \frac{E_0}{\hbar} \Im(p(t)) 
   \label{Eq8}
\end{equation} 
The set formed by eq.~\eqref{Eq7} and \eqref{Eq8} can be recast in a matrix form
\begin{equation}
   \frac{d}{dt}\begin{bmatrix} \Re(p) \\ \Im(p) \\ \Delta N \end{bmatrix} = \begin{bmatrix}  -\frac{\Delta \omega_a}{2} & 0 & 0 \\ 0 & -\frac{\Delta \omega_a}{2} & \frac{\kappa}{4 \omega_a}E_0 \\ 0 & -\frac{E_0}{\hbar} & -\frac{1}{\tau_{21}} \end{bmatrix} \begin{bmatrix} \Re(p) \\ \Im(p) \\ \Delta N \end{bmatrix} + \begin{bmatrix} 0 \\ 0 \\ \frac{N}{\tau_{21}} \end{bmatrix}
   \label{Eq9}
\end{equation} 
whose eigenvalues read 
\begin{equation}
   \left\{
    \begin{array}{ll}
         -\frac{\Delta \omega_a}{2} \\
        - \frac{1}{2}\left( \frac{\Delta \omega_a}{2}+ \frac{1}{\tau_{21}} \right)\pm \frac{1}{2} \sqrt{ \left( \frac{\Delta \omega_a}{2}- \frac{1}{\tau_{21}}\right)^2 -   \frac{\kappa E_0^2}{\hbar \omega_a} }
    \end{array}
\right.
   \label{Eq10}
\end{equation}
Hence, if the electric field intensity is large enough to fulfill
\begin{equation}
   \sqrt{\frac{\kappa}{\hbar \omega_a}} |E_0| >  \left|\frac{\Delta \omega_a}{2} - \frac{1}{\tau_{21}}\right| 
   \label{Eq11}
\end{equation}
the polarization envelope and the population inversion oscillate synchronously at pulsation
\begin{equation}
   \Omega_R^{NL} = \frac{1}{2} \sqrt{ \frac{\kappa E_0^2}{\hbar \omega_a} - \left( \frac{\Delta \omega_a}{2} - \frac{1}{\tau_{21}}\right)^2}
   \label{Eq12}
\end{equation}
For large field intensity, the contribution of relaxation rate of the polarization $\Delta \omega_a$ and the spontaneous decay time rate $1/\tau_{21}$ are negligible and the non-linear Rabi pulsation reads
\begin{equation}
   \Omega_R^{NL} \approx \frac{1}{2} \sqrt{ \frac{\kappa }{\hbar \omega_a}} |E_0| 
   \label{Eq13}
\end{equation}

Eq.~\eqref{Eq13} states that $\Omega_R^{NL}$ is linearly triggered by the amplitude of the electric field which can be derived from Maxwell equations.
In time interval $[99-106] \, ps$, the source (assumed monochromatic) drives the amplitude of the electric field. 
Combining eqs.~\eqref{Eq3} for a transverse magnetic field, the electric field satisfies
\begin{equation}
   \Delta E(\textbf{r},t) - \frac{ \varepsilon(\textbf{r})}{c^2}\frac{\partial^2 E(\textbf{r},t)}{\partial^2 t} = \mu_0 \frac{\partial^2 P(\textbf{r},t)}{\partial^2 t}
   \label{Eq14}
\end{equation}
where $c$ is the vacuum speed of light fulfilling $\mu_0\varepsilon_0 c^2 = 1$.
In the polarization density we neglect the contribution of the collections of two-level atoms and only consider the polarization enforced by the point source at position $\textbf{r}_{ext}$.
In the frequency domain, eq.~\eqref{Eq14} reads
\begin{align}
   \Delta E(\textbf{r},\omega) + & \frac{\varepsilon(\textbf{r})}{c^2} \omega^2 E(\textbf{r},\omega)\nonumber \\ &= - \mu_0 \omega_{ext}^2 P_{ext} \delta(\omega-\omega_{ext})\delta(\textbf{r}-\textbf{r}_{ext})
   \label{Eq15}
\end{align}
where $\omega_{ext}$ and $P_{ext}$ stand for the frequency and amplitude of the point source excitation, respectively.
The electric field can be expanded along the modes $(\Omega_i^2,|\Psi_i \rangle)$ of the passive system, which are defined as the eigenstates of eq.~\eqref{Eq15}
\begin{equation}
   \Delta |\Psi_i \rangle +  \frac{ \varepsilon(\textbf{r})}{c^2} \Omega_i^2 |\Psi_i \rangle = 0
   \label{Eq16}
\end{equation}
with the Siegert's outgoing boundary condition \cite{Siegert1939}
\begin{equation}
   \left(\frac{d}{d \textbf{r}} - i \Omega_i \textbf{r} \right)\Big|_{|\textbf{r}| \rightarrow \infty}|\Psi_i \rangle = 0
   \label{Eq16bis}
\end{equation}
Complex frequencies $\Omega_i$ read
\begin{equation}
   \Omega_i = \nu_i - i \frac{\Gamma_i}{2}
   \label{Eq17}
\end{equation}
where is $\nu_i$ stands for the frequency and $\frac{\Gamma_i}{2} > 0$ the damping of mode $i$. 
The modes are in the regime of Anderson-localization \cite{Anderson1958} and thus experiment small leakage (see Fig.~\ref{Fig1}(b)).
Therefore, we assume the modes to be almost orthogonal
\begin{equation}
   \langle \Psi_q | \varepsilon(\textbf{r}) | \Psi_p \rangle \approx \delta_{pq} 
   \label{Eq18}
\end{equation}
Expansion of the electric field along the modes reads
\begin{equation}
   E(\textbf{r},\omega) = \sum a_i(\omega) |\Psi_i \rangle 
   \label{Eq19}
\end{equation}
Inserting electric field expansion, eq.~\eqref{Eq15} reads
\begin{align}
   \sum a_i(\omega)&  \left( \Delta |\Psi_i \rangle +  \frac{ \varepsilon(\textbf{r})}{c^2} \omega^2 |\Psi_i \rangle \right) \nonumber \\ & = - \mu_0 \omega_{ext}^2 P_{ext}\delta(\omega-\omega_{ext}) \delta(\textbf{r}-    \textbf{r}_{ext})
   \label{Eq20}
\end{align}
From the definition of modes provided by eq.~\eqref{Eq16}, eq.~\eqref{Eq20} reads
\begin{align}
   \sum a_i(\omega)  \frac{ \varepsilon(\textbf{r})}{c^2}&(\omega^2 - \Omega_i^2 ) |\Psi_i \rangle\nonumber \\ &  = - \mu_0 \omega_{ext}^2 P_{ext} \delta(\omega-\omega_{ext}) \delta(\textbf{r}-\textbf{r}_{ext})
   \label{Eq21}
\end{align}
Now, eq.~\eqref{Eq21} is projected along mode $\langle \Psi_i |$ 
\begin{align}
   a_i(\omega)   (\omega^2 - \Omega_i^2 )  = - \mu_0 c^2\omega_{ext}^2 P_{ext} \delta(\omega-\omega_{ext}) \Psi_i^*(\textbf{r}_{ext})
   \label{Eq22}
\end{align}
where $\Psi_i(\textbf{r}_{ext})^*$ is the conjugate amplitude of the mode $i$ at the position of the point source.
As a result the electric field reads
\begin{equation}
   E(\textbf{r},\omega) = \sum \frac{\mu_0 c^2 \omega_{ext}^2 P_{ext}}{\Omega_i^2 - \omega^2} \Psi_i^*(\textbf{r}_{ext}) \delta(\omega-\omega_{ext})  |\Psi_i \rangle 
   \label{Eq23}
\end{equation}
The point source excites resonantly the Anderson-localized mode $M$ ($\omega_{ext} = \omega_M$) which is the only excited mode: $a_{i\ne M} = 0$.
Therefore, the field at the position of the atoms $\textbf{r}_a$ reads 
\begin{equation}
   E(\textbf{r}_a,\omega) = \frac{\mu_0 c^2 \omega_{ext}^2 P_{ext}}{\Omega_M^2 - \omega_{ext}^2} \Psi_M^*(\textbf{r}_{ext}) \Psi_M(\textbf{r}_{a}) \delta(\omega-\omega_{ext})   
   \label{Eq24}
\end{equation}
Since mode $M$ has a large quality factor $Q_M = \omega_M/\Gamma_M \gg 1$, eq.~\eqref{Eq24} reads
\begin{align}
   E(\textbf{r}_a,\omega) & \approx  i \mu_0 c^2 Q_M P_{ext} \Psi_M^*(\textbf{r}_{ext}) \Psi_M(\textbf{r}_{a}) \delta(\omega-\omega_{ext}) \nonumber \\
    & = E_0 \delta(\omega-\omega_{ext})
   \label{Eq25}
\end{align}
From eq.~\eqref{Eq25} we identify the electric field amplitude $|E_0|$ and deduce the non-linear Rabi pulsation by using eq.~\eqref{Eq13}
\begin{align}
   \Omega_R^{NL} =   \sqrt{\frac{\kappa}{4 \varepsilon_0^2 \hbar \omega_a}} Q_M P_{ext} |\Psi_M(\textbf{r}_{ext})||\Psi_M(\textbf{r}_{a})|  
   \label{Eq26}
\end{align}

\subsection{Observation condition}

The condition derived in eq.~\eqref{Eq11} is not sufficient to ensure the observation of non-linear Rabi oscillations in the time domain.
This regime requires that the spacing between the two side frequencies in the frequency domain, namely $2 \Omega_R^{NL}$, must be larger than  the spectral linewidth of the central component, namely $F_p \Delta \omega_a$ in which $F_p$ stands for the Purcell factor \cite{Purcell1946}.
Hence, the observation condition reads
\begin{align}
   2 \Omega_R^{NL} >   F_p \Delta \omega_a  
   \label{Eq27}
\end{align}
From expression of $\Omega_R^{NL}$ given by eq.~\eqref{Eq26}, this condition can be written as
\begin{align}
   \sqrt{\frac{\mu_0^2 \kappa}{\hbar \omega_a}} c^2 Q_M P_{ext} |\Psi_M(\textbf{r}_{ext})||\Psi_M(\textbf{r}_{a})|  >   F_p \Delta \omega_a  
   \label{Eq28}
\end{align}
The Purcell factor $F_p$ can be derived from the local density of states (LDOS) spectrum $\rho(\textbf{r},\omega)$ \cite{Caze2013}, which reads 
\begin{align}
   \rho(\textbf{r},\omega) = \sum_n \frac{|\Psi_n(\textbf{r})|^2}{\pi} \frac{\Gamma_n/2}{(\omega_n - \omega)^2 + (\Gamma_n/2)^2} 
   \label{Eq29}
\end{align}
where index $n$ stands for the contribution of the different modes.
In the Anderson regime of localization, the spectral overlap between modes is weak and thus the LDOS spectrum at frequency $\omega_M$ reads 
\begin{align}
   \rho(\textbf{r},\omega_M) = \frac{2 |\Psi_M(\textbf{r})|^2}{\pi \Gamma_M}  
   \label{Eq29bis}
\end{align}
For a specific position and frequency, the Purcell factor $F_p$ is defined as the ratio between LDOS spectrum and the vacuum LDOS, namely $\rho_0(\omega) = \omega/(2 \pi c^2)$.
Therefore, from eq.~\eqref{Eq29bis} we obtain
\begin{align}
   F_p = 4 c^2 \frac{|\Psi_M(\textbf{r}_a)|^2}{\omega_M \Gamma_M} 
   \label{Eq30}
\end{align}
Combining eq.~\eqref{Eq28} and \eqref{Eq30}, we obtain the observation condition for the non-linear Rabi regime 
\begin{align}
   \sqrt{\frac{ \kappa}{\varepsilon_0^2 \hbar \omega_a}} P_{ext} |\Psi_M(\textbf{r}_{ext})|  >   \frac{4 c^2}{\omega_M^2} |\Psi_M(\textbf{r}_a)| \Delta \omega_a  
   \label{Eq31}
\end{align}

\section{Theoretical investigation of the linear Rabi regime}\label{SecIV}
In this section, we describe analytically the regime of oscillations observed in time interval $[200-226] \, ps$ (see Fig.~\ref{Fig3}(c)) and referred to as the linear Rabi regime.
In a similar way to Sec.~\ref{SecIII}, we derive the pulsation of linear oscillations and the observation condition.

\subsection{Linear Rabi pulsation}
In time interval $[200-226] \, ps$ (see Fig.~\ref{Fig3}(c)), the electric field is too small to saturate the two-level atoms. 
The population inversion no longer oscillates and is almost constant.
Therefore, the collection of two-level atoms is characterized by a linear susceptibility.
The interaction between the atoms and the Anderson-localized mode results in oscillations of the polarization density and the electric field at the linear Rabi pulsation $\Omega_R^L$.
This regime is referred to as the linear Rabi regime.

As shown by Fig.~\ref{Fig3}(c), the population inversion reads $\Delta N \approx N$ for small intensity of the electric field.
From eq.~\eqref{Eq6}, we derive the linear relation between the electric field and the polarization density in the frequency domain at the position $\textbf{r}_a$ of the atoms:
\begin{equation}
   P(\textbf{r}_a,\omega) =  \frac{i \frac{\kappa}{2 \omega_a} N}{i (\omega_a  - \omega) + \frac{\Delta \omega_a}{2}} E(\textbf{r}_a,\omega) 
   \label{Eq32}
\end{equation} 
Inserting eq.~\eqref{Eq32} in the electric field equation provided by eq.~\eqref{Eq14}, we obtain 
\begin{align}
   \Delta E(\textbf{r},\omega) & + \omega^2 \frac{\varepsilon(\textbf{r})}{c^2} E(\textbf{r},\omega) \nonumber \\ &= \mu_0 \omega^2   \frac{i \frac{\kappa}{2 \omega_a} N}{i (\omega  - \omega_a) - \frac{\Delta \omega_a}{2}} E(\textbf{r}_a,\omega) \delta(\textbf{r}-\textbf{r}_a)
   \label{Eq33}
\end{align}
Using the electric field expansion along the modes given in eq.~\eqref{Eq19} and projecting along $\langle \Psi_M|$, eq.~\eqref{Eq33} reads  
\begin{align}
     (\omega^2-&\Omega_M^2)  \frac{a_M(\omega)}{c^2}  \nonumber \\ &=  \mu_0 \omega^2  \frac{i \frac{\kappa}{2 \omega_a} N}{i (\omega  - \omega_a) - \frac{\Delta \omega_a}{2}} \sum_i a_i(\omega) \Psi_M(\textbf{r}_a)^* \Psi_i(\textbf{r}_a)
   \label{Eq34}
\end{align}
Since only mode $M$ is initially excited: $a_{i\ne M} = 0$.
Therefore, eq.~\eqref{Eq34} reads:
\begin{align}
     (\omega^2-&\Omega_M^2)    =   \frac{\kappa \omega^2 N}{2 \omega_a \varepsilon_0}  \frac{| \Psi_M(\textbf{r}_a)|^2}{ (\omega  - \omega_a) + i\frac{\Delta \omega_a}{2}}  
   \label{Eq35}
\end{align}
The atomic transition is resonant with mode $M$ ($\omega_M = \omega_a$), thus using the rotating wave approximation eq.~\eqref{Eq35} can be recast in a second order polynomial form:
\begin{align}
     \left(\omega-\omega_M + i \frac{\Gamma_M}{2}\right) \left(\omega  - \omega_M + i \frac{\Delta \omega_a}{2} \right)   =   \frac{\kappa N}{4  \varepsilon_0} | \Psi_M(\textbf{r}_a)|^2  
   \label{Eq36}
\end{align}
whose solutions read
\begin{align}
   \omega_\pm = \omega_M - i \frac{\Gamma_M + \Delta \omega_a}{4} \pm \sqrt{\frac{\kappa N}{4  \varepsilon_0}| \Psi_M(\textbf{r}_a)|^2 - \frac{(\Gamma_M - \Delta \omega_a)^2}{16}}
   \label{Eq37}
\end{align}
Under the condition
\begin{align}
   \frac{\kappa N}{4  \varepsilon_0}| \Psi_M(\textbf{r}_a)|^2  \ge \frac{(\Gamma_M - \Delta \omega_a)^2}{16}
   \label{Eq38}
\end{align}
the electric field is split in two distinct frequencies components (see Fig.~\ref{Fig4}(b)).
This linear Rabi splitting corresponds to oscillations in the temporal domain of the electric field and the polarization density at pulsation:
\begin{align}
   \Omega_R^L = \frac{\omega_+ - \omega_-}{2} =  \sqrt{\frac{\kappa N}{4  \varepsilon_0}| \Psi_M(\textbf{r}_a)|^2 - \frac{(\Gamma_M - \Delta \omega_a)^2}{16}}
   \label{Eq39}
\end{align}

\subsection{Observation condition}
The condition derived in eq.~\eqref{Eq38} is not sufficient to ensure that the linear Rabi splitting is larger than the linewidth of the electric field at frequencies $\omega_\pm$.
Hence, to observe a frequency splitting and temporal oscillation, we must satisfy 
\begin{align}
   \Omega_R^L \ge \frac{\Gamma_M + \Delta \omega_a}{4}
   \label{Eq40}
\end{align}
Using the expression of the linear Rabi pulsation provided in eq.~\eqref{Eq39}, eq.~\eqref{Eq40} reads     
\begin{align}
   \frac{\kappa N}{4  \varepsilon_0}| \Psi_M(\textbf{r}_a)|^2 \ge \frac{\Gamma_M^2 + \Delta \omega_a^2}{8}
   \label{Eq41}
\end{align}
Since the damping of mode $M$ is much larger than the relaxation rate of the polarization, $\Gamma_M \gg \Delta \omega_a$, the linear Rabi regime condition reads  
\begin{align}
   \sqrt{\frac{2 \kappa N}{ \varepsilon_0}} | \Psi_M(\textbf{r}_a)| \ge \Gamma_M
   \label{Eq42}
\end{align}

\section{Successive non-linear and linear Rabi oscillations in the transient regime}
\label{SecV}
In this section, we investigate the transition from non-linear to linear Rabi regimes observed numerically in Sec.~\ref{SecII}, when the external source is turned off.
First, we theoretically derive the linear and non-linear Rabi pulsations for a decaying electric field amplitude.
Theoretical predictions are consistent with FDTD simulations presented in Sec.~\ref{SecII}.
Then we explain the transition from non-linear to linear Rabi regime observed in time interval $[143-174] \, ps$. 

\subsection{Transient regime}
\label{SecV1}
We consider the system after the external source has been turned off at $t_e \sim 120 $ ps.
In this transient regime, the amplitude of the electric field decays as $e^{-\Gamma_M t/2}$. 
We know from eq.~\eqref{Eq13} that the amplitude of the electric field drives the amplitude of the non-linear Rabi pulsation $\Omega_{R}^{NL}$.
Therefore, we can derive from eq.~\eqref{Eq26} the time evolution of $\Omega_{R}^{NL}$ in this transient regime:
\begin{align} 
   \Omega_R^{NL}(t) = \sqrt{ \frac{\kappa }{4 \varepsilon_0^2 \hbar \omega_a}} Q_M P_{ext} | \Psi_M(\textbf{r}_{ext})| | \Psi_M(\textbf{r}_a)| e^{-\frac{\Gamma_M}{2} (t-t_e)} 
   \label{Eq43}
\end{align}
In contrast, the linear Rabi regime results from a linear coupling between the Anderson-localized mode and the atoms.
As illustrated in eq.~\eqref{Eq39}, the amplitude of the linear Rabi pulsation $\Omega_{R}^{L}$ is independent from the amplitude of the electric field and therefore remains unchanged during the transient regime.

\begin{figure}[h!]
   {\includegraphics[scale=.585]{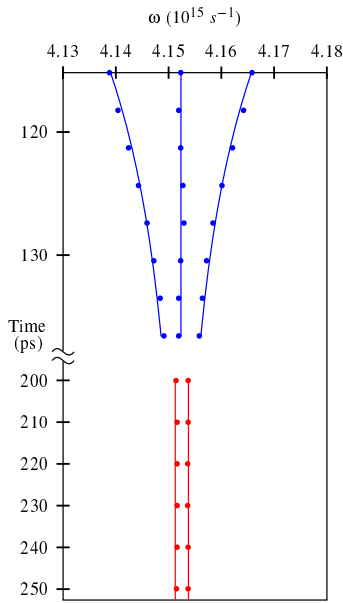}} 
\caption{ 
Evolution of the frequency components of the electric field and the polarization over time:
Blue dots represent the position of the triplet in the non-linear Rabi regime computed from FDTD simulation in the time interval $[115-135]$ $ps$.
The blue lines correspond to the temporal evolution of $\Omega_R^{NL}(t)$ predicted by eq.~\eqref{Eq43}.
Red dots are the position of the linear Rabi splitting obtained from FDTD simulations in the time interval $[200-255]$ $ps$.
Red lines stand for the prediction of the linear splitting $\Omega_R^L$ predicted by eq.~\eqref{Eq38}.
} \label{Fig7}
\end{figure}

To confirm these predictions, we use a Finite Element Method (FEM) to compute the modes of the disordered system presented in Fig.~\ref{Fig1}(a).
In particular, we obtain the Anderson-localized mode $M$: $(\Omega_M^2,\Psi_M(\textbf{r}))$.  
We first consider the time interval $[115-135]$ $ps$.
In this interval, the system evolves in the non-linear Rabi regime and the electric field exhibits three components forming a triplet (see Fig.~\ref{Fig4}(a)).
In Fig.~\ref{Fig7}, we plot the positions of the triplet observed in FDTD simulation (blue dots) over time.
Using FEM computation of mode $M$, we predict the spectral position of the triplet versus time (blue line), which is shown in Fig.~\ref{Fig7} and compared to the FDTD simulations.
The theoretical prediction provided by eq.~\eqref{Eq43} turns to fit nicely with our numerical simulations.
Now, we consider the time interval $[200-250]$ $ps$, where a linear Rabi splitting is observed.
We plot both the measured positions of the two peaks over time (red dots) and the theoretical splitting provided by eq.~\eqref{Eq39} (red line).
As predicted the linear Rabi splitting is independent of the electric field amplitude.
Moreover, the measured linear Rabi splitting ($\Omega_R^L \approx 0.6\, 10^{12} \, s^{-1}$) is consistent with the prediction of eq.~\eqref{Eq39} ($\Omega_R^L = 0.67\, 10^{12} \, s^{-1}$).  

\floatsetup[figure]{style=plain,subcapbesideposition=top}

\subsection{Transition from non-linear to linear regime}
The evolutions of non-linear and linear pulsations for a decaying electric field, derived in Sec.~\ref{SecV1}, correspond to two limits.
In the non-linear regime the contribution of linear oscillations is totally neglected and \textit{vice versa}.
In reality, they coexist, especially in the transient regime.
Nevertheless, these two limits nicely explain the progressive transition from non-linear to linear regime observed in Fig.~\ref{Fig3}(b).

When the amplitude of the electric field is large ($\Omega_R^{NL} (t) \gg \Omega_R^{L}$) the appearance of linear Rabi oscillations is hindered by non-linear oscillations (see  Fig.~\ref{Fig3}(a)).
The progressive decay of the field amplitude leads to a decrease of the non-linear splitting.
When $\Omega_R^{NL} (t) \sim \Omega_R^{L}$ both effects coexist.
In Fig.~\ref{Fig3}(b), the population inversion and the polarization oscillate synchronously (non-linear regime) while the electric field oscillates in quadrature with the polarization (linear regime).
When $\Omega_R^{NL} (t) < \Omega_R^{L}$, the linear Rabi pulsation prevails over the non-linear one and the oscillations of the polarization envelope simultaneously slow down (at $t_{trans} \approx 160$ $ps$ in Fig.~\ref{Fig3}(b)). 
Finally, when the field intensity becomes too small to saturate the population inversion (see Fig.~\ref{Fig3}(c)), the non-linear oscillations disappear ($\Omega_R^{NL} (t) \ll \Omega_R^{L}$) and only the linear regime remains. 

Asymptotic expressions of $\Omega_R^{NL} (t)$ and $\Omega_R^{L}$ allow to go further and to estimate the transition time between both regimes by writing
\begin{align} 
   \Omega_R^{NL}(t_{tran}) = \Omega_R^{L}  
   \label{Eq44}
\end{align}
where $t_{tran}$ stands for the transition time.
The computed value $t_{tran} \approx 163$ $ps$ is in good agreement with the transition time observed in FDTD simulation at $t_{trans} \approx 160$ $ps$ (see  Fig.~\ref{Fig3}(b)).

\subsection{Observation condition of both regimes}
In Sec.~\ref{SecIII} and \ref{SecIV} we derived successively the observation conditions for non-linear and linear oscillations given by eq.~\eqref{Eq31} and \eqref{Eq42}, respectively.
Both relations stress the influence of the coupling between the Anderson-localized mode and the atoms, namely $\Psi_M(\textbf{r}_a)$.

In the non-linear condition provided by eq.~\eqref{Eq31}, $\Psi_M(\textbf{r}_a)$ is responsible for the spectral linewidth of the central component of the electric field.
Hence, to reach the non-linear regime with a small external excitation, $P_{ext}$, $\Psi_M(\textbf{r}_a)$ must be chosen as small as possible.
In the linear regime, energy is reversibly exchanged between the mode and the emitter.
Therefore a large coupling between the mode and the emitter is required to fulfill eq.~\eqref{Eq42}.

Remarkably, to achieve the successive observation of non-linear and linear oscillations, the coupling between the emitter and the mode, $\Psi_M(\textbf{r}_a)$, must be carefully chosen.
The condition to observe linear and non-linear regimes in steady state can be derived from eq.~\eqref{Eq31} and \eqref{Eq42} and reads 
\begin{align}
  \Gamma_M \sqrt{\frac{\varepsilon_0}{2 \kappa N}} \le | \Psi_M(\textbf{r}_a)| \le \frac{\mu_0 }{4} \sqrt{\frac{\kappa \omega_M^3}{\hbar}} \frac{P_{ext}}{\Delta \omega_a} | \Psi_M(\textbf{r}_{ext})|
   \label{Eq49}
\end{align}
where the left (right) hand side of eq.~\eqref{Eq49} must be fulfilled at small (large) electric field intensity.

\section{Conclusion}
\label{SecVI}
In this article, we studied the dynamic interaction of a collection of two-level atoms and an Anderson-localized mode.
We demonstrated that such interaction results in two kinds of Rabi oscillations of the electric field.
These two different regimes, referred to as linear and non-linear Rabi regimes, are due to two distinct physical mechanisms.
The linear regime is reached at small electric field intensity when the collection of atoms strongly couples to the Anderson-localized mode \cite{Sapienza2010,Thyrrestrup2012,Caze2013}.
This semiclassical regime of strong coupling is referred to as the NPNMC in the literature \cite{Khitrova2006} and results in a splitting of the atomic frequency.
The non-linear regime is observed at larger intensity.
In this regime, the electric field saturates the population inversion which starts to oscillate.
These oscillations are responsible for new transitions in the atomic energy structure resulting in a non-linear atomic susceptibility.
Such a susceptibility induces non-linear spectral components of the field which translate into Rabi oscillations in the time domain.
This physical effect is known as the Dynamic Stark Effect splitting \cite{Meystre2007,Boyd2008}.
It is predicted here in the framework of disordered media and Anderson-localized modes.  
For a progressively decaying field intensity, we qualitatively and quantitatively explained the transition from one regime of oscillations to the other.
Moreover we emphasized that coupling between the mode and the collection of atoms must be adequately chosen to ensure the observation of both regimes.

In this work, we confirmed the ability of 2D random media to support NPNMC \cite{Caze2013}, which offers promising applications for the development of disorder robust cQED \cite{Sapienza2010}.
Remarkably, we also demonstrated the capacity of Anderson-localized modes to achieve strong and non-linear light-matter interaction characterized by DSE splitting.
In terms of perspectives, this work opens further the way to various applications.
For instance, sidebands resulting from the DSE were proposed to achieve coherent amplification \cite{Guerin2008} and could be used to achieve new random laser sources \cite{Froufe-Perez2009,Guerin2010}.
Anderson-localized modes can also be externally tuned by local modifications of the medium \cite{Bachelard2014a, Riboli2014}.
Hence, one can perform a control of the non-linear atomic emission similar to what was achieved recently in disordered media with a different gain mechanism \cite{Bachelard2012,Leonetti2013,Bachelard2014}.   
Moreover, the DSE is known to be closely related to the phenomenon of resonance fluorescence \cite{Mollow1969}.
Therefore, we believe that Anderson-localized modes may serve to induce resonance fluorescence and thus design cheap and robust quantum light sources (e.g. single photon source \cite{Flagg2009,Ulhaq2012}).
In summary, we believe that this work opens further the way to the achievement of non-linear optics (e.g. lasing without inversion or wave-mixing) in strongly disordered media.

\section*{Acknowledgement}
We acknowledge A. Goetschy and R. Pierrat for fruitfull discussions. 
PS thanks the Agence Nationale de la Recherche for its support under grant ANR PLATON (No. 12-BS09-003-01). 
PS and RC are thankful to the LABEX WIFI (Laboratory of Excellence within the French Program "Investments for the Future") under references ANR-10-LABX-24 and ANR-10-
IDEX-0001-02 PSL* and the Groupement de Recherche 3219 MesoImage for their financial support.
\newline

\end{document}